\documentclass[%
 reprint,
superscriptaddress,
nofootinbib,
 amsmath,amssymb,
 aps,
prd,
]{revtex4-2}

\usepackage{graphicx}
\usepackage{dcolumn}
\usepackage{bm}
\usepackage{lipsum}
\usepackage{xfrac}

\usepackage{hyperref}

\begin{document}


\title{Towards a neutrino-limited dark matter search with crystalline xenon}

\author{H. Chen}\email{maque@lbl.gov}
\affiliation{Lawrence Berkeley National Laboratory, 1 Cyclotron Rd, Berkeley, CA 94720, USA}
\author{R. Gibbons}
\affiliation{Lawrence Berkeley National Laboratory, 1 Cyclotron Rd, Berkeley, CA 94720, USA}
\affiliation{University of California, Berkeley, Department of Physics, Berkeley, CA 94720, USA}
\author{S.J. Haselschwardt}
\affiliation{Lawrence Berkeley National Laboratory, 1 Cyclotron Rd, Berkeley, CA 94720, USA}
\author{S. Kravitz}
\affiliation{The University of Texas at Austin, Department of Physics, 2515 Speedway, Austin, Tx 78712}
\author{Q. Xia}%
\affiliation{Lawrence Berkeley National Laboratory, 1 Cyclotron Rd, Berkeley, CA 94720, USA}
\author{P. Sorensen}\email{pfsorensen@lbl.gov}
\affiliation{Lawrence Berkeley National Laboratory, 1 Cyclotron Rd, Berkeley, CA 94720, USA}

\date{\today}

\begin{abstract}
Experiments searching for weakly interacting massive particle dark matter are now detecting background events from solar neutrino-electron scattering. However, the dominant background in state-of-the-art experiments such as LZ and XENONnT is beta decays from radon contamination. In spite of careful detector material screening, radon progenitor atoms are ubiquitous and long-lived, and radon is extremely soluble in liquid xenon. We propose a change of phase and demonstrate that crystalline xenon offers more than a factor $\times500$ exclusion against radon ingress, compared with the liquid state. 
This level of radon exclusion would allow crystallized versions of existing experiments to probe spin-independent cross sections near $10^{-47}$~cm$^2$ in roughly 11~years, as opposed to the 35~years required otherwise. 
\end{abstract}

\maketitle


\section{\label{sec:level1}Introduction}
Astrophysical observations and cosmological models suggest that 84\% of total matter density in the universe is non-luminous, non-baryonic dark matter \cite{Planck:2018vyg, Sofue:2000jx}. The majority, if not all, of the dark matter requires new particles beyond the Standard Model \cite{Workman:2022ynf}. For decades, the most well-motivated model has been the Weakly Interacting Massive Particle (WIMP) \cite{BAUDIS201294}. 
Significant experimental effort has been dedicated to detecting WIMP-nucleon interactions \cite{Akerib:2022ort}, but unambiguous evidence has yet to be observed \cite{Schumann:2019eaa}.  

The most sensitive exclusion limits on the interaction cross section for WIMPs with masses greater than about 10 GeV are due to the  LUX-ZEPLIN (LZ) experiment \cite{LUX-ZEPLIN:2022xrq}, which utilizes a 7~tonne active target comprised of a liquid xenon time projection chamber (TPC). The LZ background rate in the approximately $1-10$~keV energy range of interest for WIMP scattering is $<$ 1 mHz/kg/day \cite{LUX-ZEPLIN:2022psu}. About $\sfrac{1}{10}$ of these events are due to irreducible solar neutrino-electron scattering. Eventually, it is hoped that this class of experiments will reach a sensitivity that is limited by coherent neutrino-nucleus scattering of atmospheric neutrinos \cite{Akerib:2022ort}, which is variously referred to as the neutrino detection limit, the neutrino floor, or the neutrino fog~\cite{OHare:2021utq}. However, progress towards this goal is obscured by the fact that at present, some $\sfrac{2}{3}$ of the observed background events are due to ground-state beta decays of $^{214}$Pb and $^{212}$Pb, from the decay of $^{222}$Rn and $^{220}$Rn~\cite{LUX-ZEPLIN:2022xrq}. 

Radon is ubiquitous due to the long half lives of its uranium and thorium progenitors. It is typically present at ppm levels \cite{Formaggio:2004ge} and emanates via diffusion from a wide range of materials. It is generally extremely soluble in liquids, including liquid xenon. Pre-screening and careful selection of materials prior to detector construction is essential but cannot remove all the radon.  LZ employs a charcoal chromatography radon reduction system \cite{LZ:2020fty} in the vapor phase for on-line purification. XENONnT \cite{XENONCollaboration:2023orw} has achieved a somewhat lower level of radon background using an in-line cryogenic distillation column \cite{Murra:2022mlr}. An intrinsic limitation to these radon reduction techniques is that they must compete against continuous emanation of radon from detector materials.  This challenge has led us to consider crystalline xenon as a detector medium. The motivation is that it might prevent radon which emanates from detector materials from mixing into the xenon target, and that it might allow one to tag the decays of radon daughters trapped in the crystal matrix.

In a previous article \cite{Kravitz:2022mby} we demonstrated that crystalline and liquid xenon have nearly identical scintillation yields, and similar ionization yields. The higher electron mobility in crystalline xenon observed both in our work and a previous study \cite{PhysRevB.10.4464} is an additional benefit for the suppression of pile-up events. An earlier study showed that ionized electrons are more easily emitted into the vapor above crystalline xenon, compared with liquid xenon \cite{gushchin:1982emi}. In this article, we demonstrate the key motivation for this technology: crystalline xenon's ability to exclude radon contamination. We also offer a hint at crystalline xenon's potential to tag the nuclear decay chain of radon progeny. The effect on future dark matter searches is quantified.

\section{Radon exclusion from crystalline xenon}
In order to quantify the transport of radon across the xenon liquid/vapor and xenon crystal/vapor interface, we used a dual-phase TPC previously described in Ref.~\cite{Kravitz:2022mby}. We first used a low-activity $\mathcal{O}(1)$~Hz source of $^{222}$Rn ($t_{1/2} = 3.8~$days) operated over a period of about a month. We then followed up with several short experiments utilizing a higher-activity $\mathcal{O}(100)$~Hz source of $^{220}$Rn ($t_{1/2} = 56~$seconds) to improve the sensitivity of the result. The radon isotopes were introduced into the vapor above the liquid or crystalline xenon via a sealed xenon circulation loop, which continuously purifies the xenon. Because the surface area of the liquid/vapor and crystal/vapor interfaces are constant, a simple counting experiment suffices. 

\subsection{Instrument}\label{instrument}
\begin{figure}[h]
	\centering
	\includegraphics[width=0.9 \columnwidth]{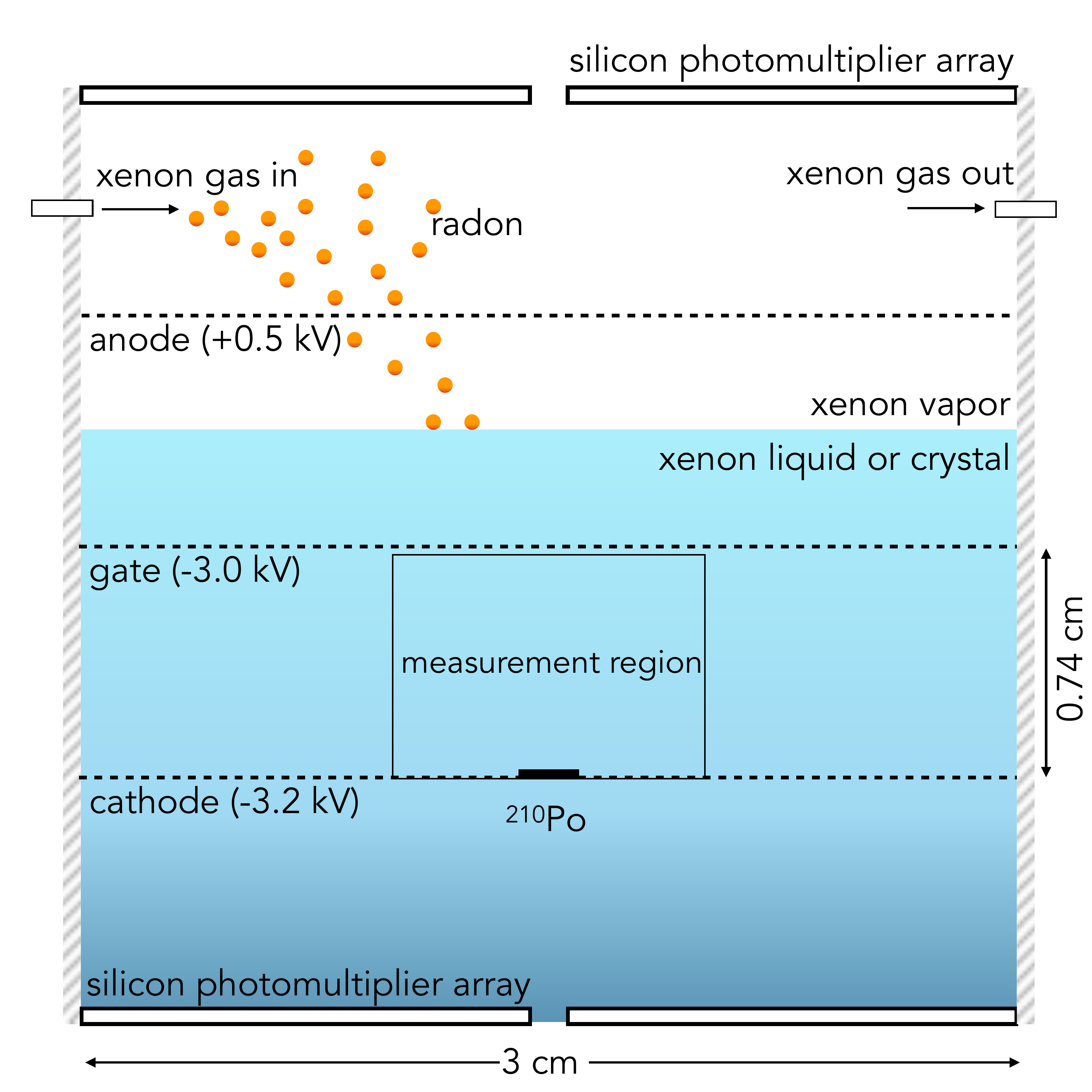}
	\caption{Schematic of the experimental apparatus, approximately to scale. Radon atoms (indicated) were introduced via the sealed xenon gas circulation loop.}
	\label{fig:0}
\end{figure}

The TPC is a cylindrical structure made of polytetrafluoroethylene with an internal radius of $1.5$~cm. A schematic is shown in Fig. \ref{fig:0}. Three electrodes (cathode, gate and anode, in order from bottom to top) are used to define the electric fields across the xenon. The distance between each electrode is 7.4~mm. The liquid/vapor (or crystal/vapor) interface was about half way between the gate and the anode \footnote{The interface height was slightly different for the liquid and crystal data. This is because the process of crystallizing the liquid xenon requires overfilling the liquid level so it is above the anode; the change in density then brings the crystal/vapor interface back between the gate and the anode. The height was not quantified. Additionally, the data reported in Sec \ref{222} and Sec. \ref{220} have slight differences in the height of the interface. This affects the S2 signal size.}. Several upgrades were made to the instrument since the previous work \cite{Kravitz:2022mby}. The present system utilizes two 16-channel arrays of silicon photomultiupliers (Hamamatsu S13371-6050CQ) for detection of 175~nm scintillation and electroluminesence photons. One array is located below the cathode, and the other is located above the anode. Also, the previous high voltage electrode grids were replaced with more robust electro-formed meshes. 

Particle interactions generate both prompt scintillation photons (referred to as S1) and ionization signals (referred to as S2). The ionized electrons are drifted across the liquid/crystal by an electric field applied between the cathode and the gate electrode, then emitted into the vapor phase above by a stronger electric field between the gate and anode. Acceleration of the electrons through the vapor produces an electroluminescence signal proportional to the number of electrons (the S2 signal). The time difference between the S1 and S2 signals corresponds to the $z$ coordinate of the interaction. The photon distribution of the S2 signal gives the $(x,y)$ coordinates.

\begin{figure}[h]
	\centering
	\includegraphics[width=1 \columnwidth]{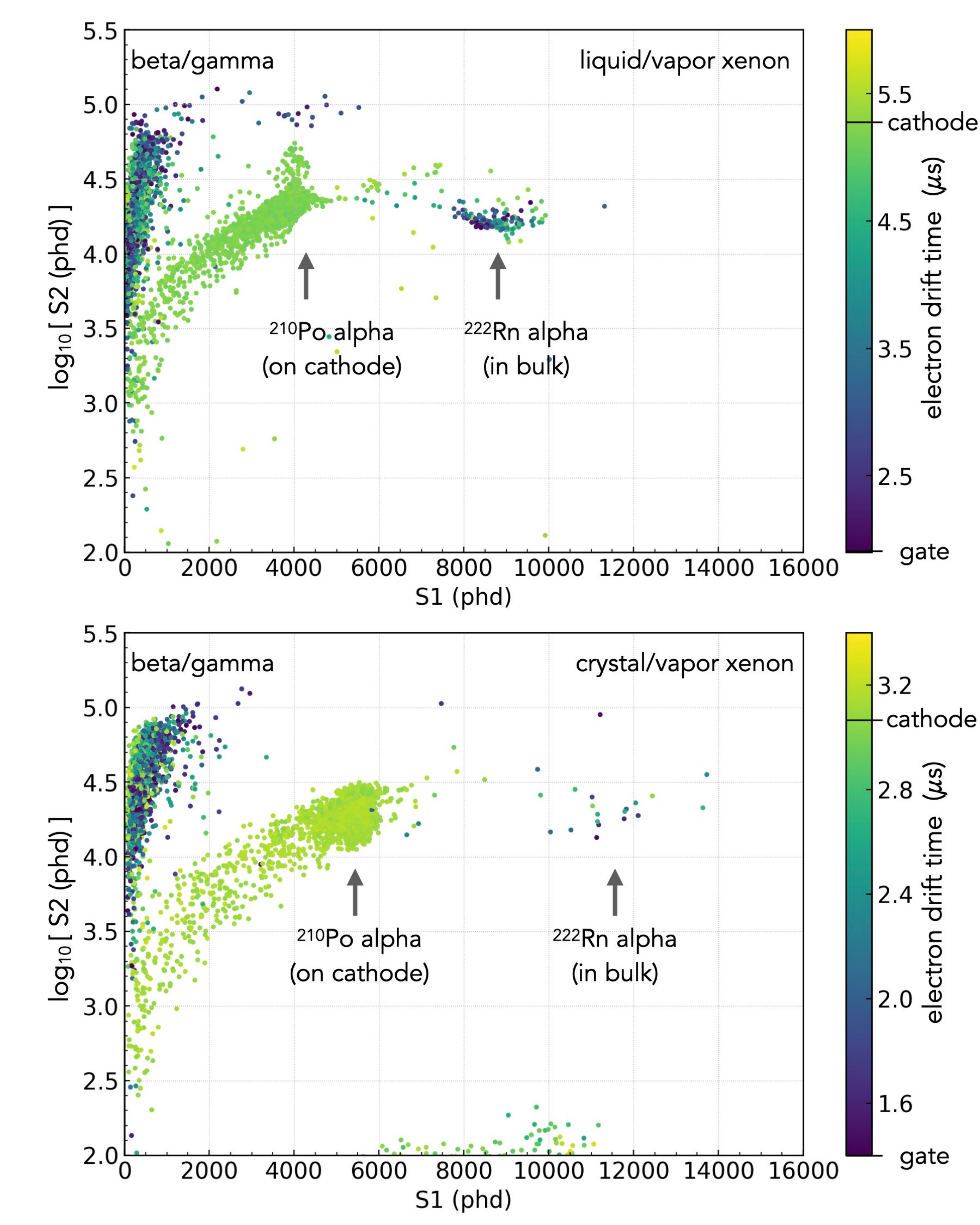}
	\caption{{\bf Top}: Distribution of scintillation (S1) and ionization (S2) signals from all single scatter events in the central region of the TPC with $r<1$~cm, after circulating $^{222}$Rn in liquid/vapor mode for 210 hours. {\bf Bottom}: The same distributions after circulating $^{222}$Rn in crystal/vapor mode for 300 hours. Residual radon events in the crystal remain from the earlier liquid phase period. Arrows in Fig.~\ref{fig:2} at $t=-110$ and $t=310$ indicate the corresponding data points. The population of events with S2~$\sim200~$phd are due to $^{210}$Po source decays at the edge of the disk.}
	\label{fig:1}
\end{figure}

\begin{figure*}
  \includegraphics[width=\textwidth]{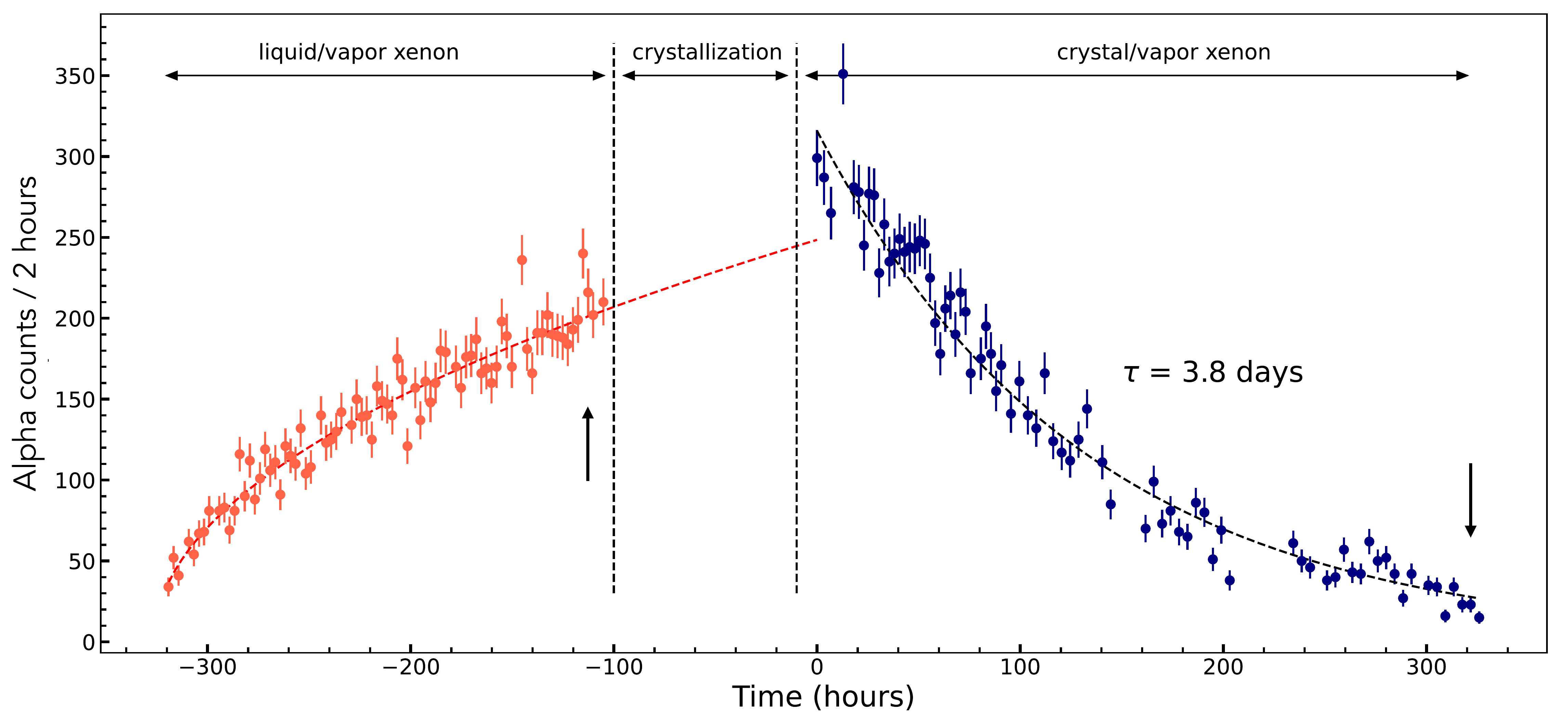}
  \caption{Observed count rate of alpha decays in the bulk xenon region defined in Sec. \ref{instrument}. Gas xenon was continuously circulated through a low-activity $^{222}$Rn source during the entire time shown. Until $t \approx -100$~hours, the TPC was operated in liquid/vapor mode and the count rate of alpha decays increased $\propto{\sqrt{t}}$ as radon diffused into the bulk liquid xenon. Following crystallization, the count rate in crystal/vapor mode decreased following the expected $\tau=3.8$~day half life of $^{222}$Rn indicating that no new radon could enter the crystal in spite of its continuously refreshed presence in the vapor. The arrows at $t\sim-110$~hours and $t\sim315$~hours indicate the two data sets shown in Fig. \ref{fig:1}.}
  \label{fig:2}
\end{figure*}

The primary results of this work require counting a number of alpha particle events from the radon decay chains, so we deposited a $^{210}$Po alpha particle calibration source in the center of the cathode grid. The source emits $5.3$~MeV alphas which is a smaller energy than that of the various alpha decays we need to count. For the data reported in Sec \ref{222}, the source was deposited on a 3~mm diameter integrated disk. For the subsequent data reported in Sec. \ref{220} we removed the disk and deposited the source directly on the cathode mesh. This change leads to an increase in the photon detection efficiency and a wider dispersion of the source response, due to the trajectories of alpha particles with respect to the mesh and to the electric field lines. In both cases, the alpha population is easy to identify by the reconstruction of deposited energy and the location of the event. We used the alpha rates detected from $^{210}$Po in liquid/vapor mode and crystal/vapor mode to confirm that the detector had similar detection efficiency for MeV alphas in each of the two operation modes. During the experiments, verification of the stability of the detector response was monitored by detection of the photo-absorption peak from 122~keV and 136~keV ($^{57}$Co) gammas.

The TPC was operated with the cathode at -3.2 kV, the gate at -3.0 kV and the anode at +0.5 kV. To avoid non-uniformity (fringing) of the applied electric fields, valid single scatter events were required to occur within a central cylinder defined by $r<1$~cm, where $r$ is the distance from the symmetry axis of the TPC. The active region of the TPC between the cathode and the gate electrodes extends 7.4~mm in $z$. An additional $3-4$~mm of active xenon liquid or crystal was present above the gate electrode (this value was slightly different in each case). Events in the vapor are rejected based on their drift time, and/or by a reduced S2 size.  An example of data satisfying these selection criteria in both the liquid and crystal phases are shown in Fig. \ref{fig:1}. Beta and gamma background events are well-separated from the alpha populations. As with $^{210}$Po alpha decays from the cathode, bulk alpha decays are also easily selected by their energy and location of the event.

The trigger for these data was a simple edge threshold operated at $n\geq 2$ coincidence set at a level that easily caught alpha particle scintillation with full efficiency. The TPC was enclosed in 5~cm of lead bricks and the total background trigger rate was about 7 Hz.  The data acquisition system would saturate above 30~Hz, and the highest trigger rate for data described here was less than 20~Hz.

\begin{figure*}[t]
	\centering
	\includegraphics[width=\textwidth]{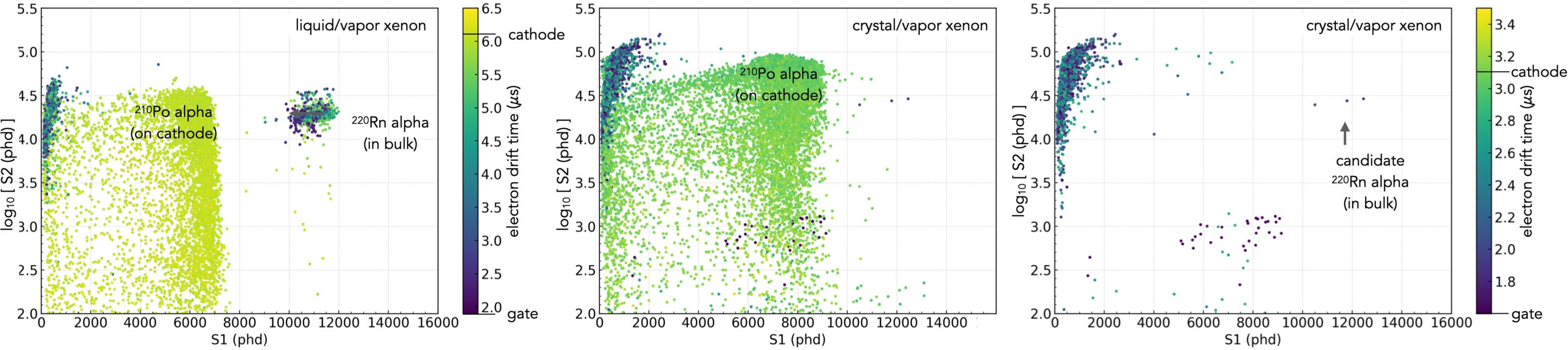}
	\caption{{\bf Left}: Distribution of scintillation (S1) and ionization (S2) signals from all single scatter events in the central region of the TPC obtained during four hours of continuously circulating $^{220}$Rn in liquid/vapor mode. {\bf Center}: The same distributions obtained during four hours of continuously circulating $^{220}$Rn in crystal/vapor mode. {\bf Right}:  Same as Center plot, but with cathode events removed for clarity. Three candidate alpha events in the crystal are observed, at least two of which appear to be due to $^{222}$Rn rather than $^{220}$Rn. The population with S2~$\sim1000$ are due to radon decays above the anode (short electron drift times) and to the tail of the $^{210}$Po calibration source distribution (long electron drift times).}
	\label{fig:3}
\end{figure*}

\subsection{Radon exclusion measurement with $^{222}$Rn}
\label{222}
A flow-through source of $^{222}$Rn was connected in-line with the xenon gas circulation system. Xenon gas was circulated continuously at a fixed flow rate of $0.3\pm0.01$ standard liters per minute and therefore the number density of new $^{222}$Rn atoms in the vapor phase remained constant. 

Several early progeny of $^{222}$Rn, including $^{218}$Po and $^{214}$Po, emit high energy alpha particles. These are quite easy to identify by their $>5$~MeV energy, which is significantly above the beta and gamma background. The problematic background from the beta decay of  $^{214}$Pb is bracketed by these two polonium decays. 

The flow-through radon source was operated continuously and the observed count rate of alpha particles was recorded for nearly one month, as shown in Fig. \ref{fig:2}. The instrument was in liquid/vapor mode at a vapor pressure p~$\approx1.25$~Bar for the first nine days. As will be explained in Sec. \ref{220}, we found that radon dissolves from the vapor into the liquid and reaches equilibrium in less than 10 minutes; the slow increase in the alpha count rate in proportion to $\sqrt{t}$ over those nine days reflects the approach to secular equilibrium of the decay chain below $^{222}$Rn. 

Over a period of about four days, the xenon was crystallized. We did not acquire data during this time, because the crystallization procedure requires overfilling the level of liquid xenon in the instrument (which renders it impossible to observe the S2 signal). Upon completion of crystallization, we resumed the data acquisition. The count rate of alpha particles in crystal/vapor mode at a vapor pressure p~$\approx0.79$~Bar was observed for about thirteen days.  Note that neither Fig. \ref{fig:1} nor Fig. \ref{fig:2} show events from radon decay in the vapor phase, as these were easily removed by the drift time between S1 and S2.

Following crystallization of the xenon, the alpha particle count rate decreased exponentially following the 3.8~day half life of $^{222}$Rn. Because we were still continuously circulating radon gas in the vapor above the crystal, one can immediately conclude that the observed count rate was due to radon atoms that were trapped in the crystal during crystallization, and that no appreciable number of radon atoms were able to diffuse from the vapor into the crystal.

The offset between the extrapolation of liquid phase alpha counts shown in Fig. \ref{fig:2} and the start of crystal alpha counts is due to the decay of polonium daughters:  half of the $^{218}$Po and 76\% of $^{214}$Bi are left with net positive charge \cite{EXO-200:2015ura}, and in liquid can drift to the cathode where in this experiment we would not count them. In the crystalline state, the daughters are frozen in the crystal matrix and so are always counted.

Extrapolation of the liquid phase count rate trend to the end of the experiment at $t=320$~hours suggests that we should have observed about 350 counts per two~hours had we not crystallized the xenon. Given that we actually observed about 20 counts per two hours, crystallization appears to reduce the radon ingress by a factor of about $\times17.5$. It turns out this is a significant underestimate.

\subsection{Radon exclusion measurement with $^{220}$Rn}
\label{220}
In order to obtain better sensitivity in this measurement, we repeated the test procedure with a higher-activity flow-through $^{220}$Rn source, also connected in-line with the xenon gas circulation system. $^{220}$Rn and its early progeny $^{216}$Po both emit high energy alpha particles. In this case, the immediate increase in the trigger rate was a factor of about $\times2$, still dominated by alpha decays in the vapor. Data sets were started 10 minutes after initiating the radon flow, which was sufficient to saturate the observed radon event rate.

The half life of $^{220}$Rn is only 56~s, so in contrast with the previous experiment with $^{222}$Rn, we expected to count zero events from $^{220}$Rn alpha decay in crystal/vapor mode. The signal distributions are shown in Fig. \ref{fig:3}. As mentioned in Sec. \ref{instrument},the polonium source in this experiment showed a wide dispersion in response due to the trajectories of alphas. Nevertheless, it establishes the energy scale for the alpha particle events of interest. Four hour data sets recorded 2034 radon alpha events in the instrument's fiducial volume in liquid/vapor mode, compared with three events in the crystal/vapor mode. 
All three observed alpha events are nearly in the center of the crystal, about 10~mm below the crystal surface. Two of the three observed alpha events in the crystal xenon appear not from $^{220}$Rn entering the crystal, but instead due to the ubiquitous and longer-lived $^{222}$Rn, trapped in the bulk during the crystallization process. 

These events satisfy the criteria for tagging the decay of $^{222}$Rn followed by its daughter, $^{218}$Po: they occurred at the same reconstructed $(x,y,z)$ location in the crystal (within the resolution of the instrument), and with a time delay very close to the 3.1~minute half life of $^{218}$Po. The resolution of the reconstructed $z$ position is $\Delta z \sim 0.05~$mm, so this interpretation as a tagged decay sequence places an upper limit on the $z$ component of the radon daughter velocity in the crystal at $v_z \lesssim 0.05~$mm/3.1 minutes. At this rate, traversing 10~mm of crystal would require about 10 hours, and could not have been observed during this four hour data set. 

We therefore consider the 90\% confidence level Poisson upper limit on one observed events ($n=3.9$), and obtain a radon exclusion factor of more than $\times500$ at 90\% C.L. for crystalline xenon with respect to liquid xenon.  We suspect that this factor is a lower limit, but data acquisition rate limitations preclude us from testing a higher radon activity. We also note that if we were to count all three events as ingress of $^{220}$Rn, the 90\% confidence level Poisson upper limit would be $n=6.7$ and the reduction factor would be closer to $\times300$. This would still be more than sufficient to render radon irrelevant as a background for the projections we make in the next Section.

\section{Discussion}
We measured radon transport efficiency across a xenon vapor/liquid interface relative to a xenon vapor/crystal interface. We assume a similar factor would apply to other materials used for detector construction, for example $m_x$/liquid xenon vs $m_x$/crystal xenon, where $m_x$ is Teflon, titanium, stainless steel, copper, Kapton, etc. We do not expect any appreciable difference, but note it for completeness. It could be verified in future experiments.

The LZ experiment observed 182 background events from the radon chain during its first science run \cite{LUX-ZEPLIN:2022psu}. Our work suggests that if it had been able to operate in crystal/vapor mode with similar dimension and background, this could have been reduced to $<1$ background event. 

In order to quantify the benefit of a crystalline/vapor xenon TPC, we compare the projected sensitivity of an LZ-like experiment with and without radon. We use the projected background rate of 6.2 counts for 1000 days from Ref. \cite{LZ:2018qzl}, which was estimated in the energy range of 6–30 keV for nuclear recoil signals. We also assume effectively all of the radioactive krypton could be removed by e.g. charcoal chromatography, which is justified by existing technology \cite{Ames:2023lvd}. This leaves as expected background rate of 2 counts in 1000 days. This rate is dominated by solar neutrino scattering from atomic electrons, followed by two-neutrino double beta decay of $^{136}$Xe. The expected background rate for the same volume of a crystal/vapor LZ-like instrument (colloquially: ``CrystaLiZe'') would be 20\% larger due to the change in density. We assume the same discrimination of 99.5\% against electron recoils and 50\% nuclear recoil acceptance in both liquid and crystalline xenon, as in \cite{LZ:2018qzl}. 

\begin{figure}[ht]
	\centering
	\includegraphics[width=1 \columnwidth]{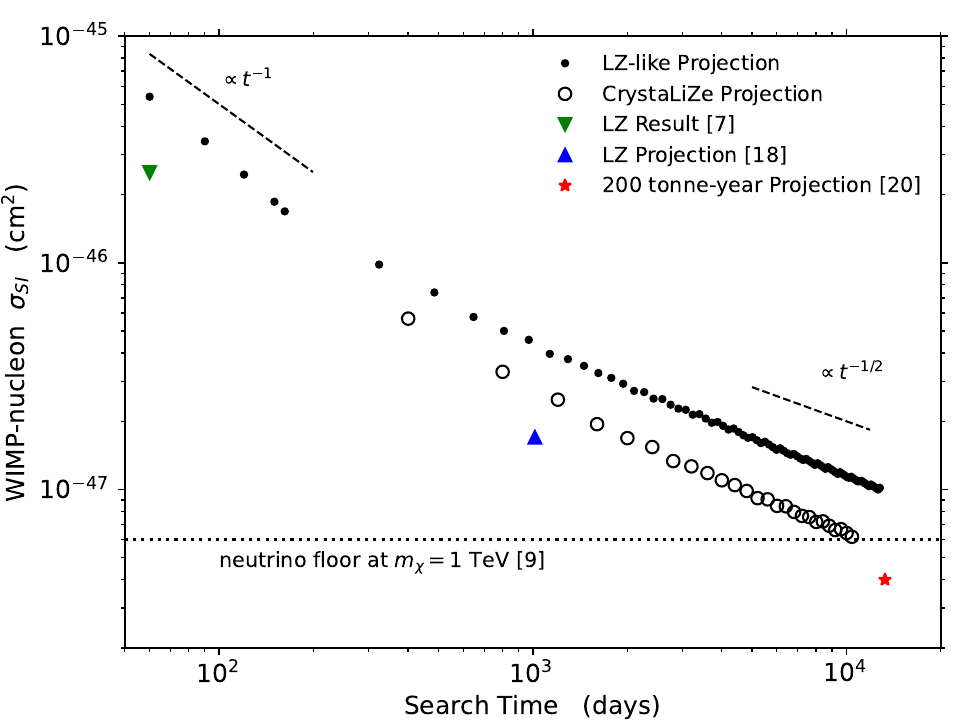}
	\caption{The projected sensitivity of an LZ-like detector to spin-independent scattering of 1~TeV dark matter particles, in the case of either 5.5 tonne active liquid xenon mass (filled circles) or 6.6 tonne active crystal xenon mass (open circles). Details of the projections are explained in the text. Also shown are the recent LZ first results \cite{LUX-ZEPLIN:2022xrq}, LZ projection \cite{LZ:2018qzl} and the projected sensitivity of a 200 tonne-year exposure of a next generation experiment \cite{Aalbers:2022dzr} scaled to a 5.5~tonne target. } 
	\label{fig:4}
\end{figure}

In Fig. \ref{fig:4}, we show how the sensitivity to the spin-independent WIMP-nucleon cross section for hypothetical WIMP mass of 1~TeV improves as a function of search time. A Feldman-Cousins ``cut-and-count'' method \cite{Feldman:1997qc} was used in both cases. The difference is significant: in order to reach a sensitivity of $\sigma=1\times10^{-47}$~cm$^2$, a radon-free crystalline xenon instrument would require about 4000 live days (11 years), a factor $\times3$ less search time than the 12700 live days (35 years) required otherwise. The former time period is likely comparable to the construction time for a next-generation instrument \cite{Aalbers:2022dzr}, while the latter is comparable to the duration of any one researcher's career. The cross section benchmark $\sigma=1\times10^{-47}$~cm$^2$ lies just above the neutrino ``floor'' as defined in Ref. \cite{OHare:2021utq}.

Liquid xenon instruments have also been used for other new physics searches beyond WIMP dark matter. One example is axion-like particle and hidden photon models, or searches for a neutrino magnetic moment \cite{XENON:2022ltv, LZ:2023poo}. In these cases, the expected signal is an electron recoil in a xenon TPC. Since these would look similar to the low-energy beta background from radon, a crystalline xenon TPC would be even more beneficial to such searches. Additionally, a proposal to dope a light element such as hydrogen or helium into a large liquid xenon TPC (HydroX) in order to increase sensitivity to low-mass dark matter \cite{Lippincott:2017yst,Haselschwardt:2023iqn} could be augmented by using crystalline xenon. This is because two key concerns for HydroX are (1) that light elements could diffuse through the seal on the photomultiplier tubes, leading to their rapid aging and demise; and (2) hydrogen may quench the S2 signal. If the light element were frozen in the crystal, these concerns could be avoided. Finally, experiments searching for the zero-neutrino mode of double beta decay in the isotope $^{136}$Xe could possibly benefit by considering crystalline xenon in the context of barium-tagging \cite{Chambers2019}.  

\section{Conclusion}
A crystal/vapor dual-phase TPC appears to be a promising detector technology for reaching the dark matter neutrino detection limit in a reasonable time scale utilizing existing experimental capabilities. Specifically, if it were found to be feasible to crystallize the LZ experiment or the XENONnT experiment following conclusion of their run plan and science goals, either of these $\mathcal{O}(10)$~tonne xenon target instruments could optimistically reach $\sigma=1\times10^{-47}$~cm$^2$ in 11 years of search time. More R\&D is needed. In particular, (a) the incident particle type discrimination of a crystal/vapor xenon TPC needs to be measured, (b) the scaling to many kilograms of target mass and beyond must be demonstrated and (c) the stability over long time periods needs to be assessed. 

\begin{acknowledgments}
Aaron Manalaysay provided advice on the Feldman-Cousins sensitivity projections. This material is based upon work supported by the U.S. Department of Energy, Office of Science, Office of High Energy Physics, under award number DE-AC02-05CH1123.
\end{acknowledgments}

\bibliography{crystalize}

\end{document}